# The notion of validity in experimental crowd dynamics

Milad Haghani

School of Civil and Environmental Engineering, The University of New South Wales, UNSW Sydney, Australia

*Corresponding author: milad.haghani@unsw.edu.au

**Abstract**

As experimental methods become increasingly popular in crowd dynamics, the question of experimental validity becomes more critical than ever. For the purposes of evaluation and interpretation of crowd experiments, it is paramount that we distinguish between different dimensions of their validity. Here, we differentiate between four types of experimentation in crowd dynamics based on their underlying *purpose of investigation*. This includes experiments whose main aim is (i) testing a behavioural theory/hypothesis, (ii) exploring empirical behavioural regularities; (iii) assessing collective (system-level) performance rather than individual behaviour (iv) calibrating and/or validating pre-existing models (including partial/local or full/global model calibration). While delineating various dimensions of external and internal validity as they relate to crowd experiments and enumerating factors that could compromise their validity, we argue how the importance of different validity dimensions varies depending on the investigation's purpose. This could provide clarity about circumstances where artificiality of laboratory settings is more justified as opposed to experiments for which higher levels of contextual fidelity and ecological validity should be given more weight. It is critical that investigators (a) be conscious of the trade-off between controllability and realism in their experiment design and (b) identify potential factors that can jeopardise validity of their experiments and take measures to mitigate such threats. Furthermore, it is argued that inferences made from *field* data do not invariably supersede those of *experiments* on the issue of validity. Depending on the purpose of investigation, maintaining internal validity sometimes necessitates laboratory settings and justifies artificiality. Naturality of observations per se does not guarantee validity. In the absence of adequate control over the causal relationship, the question of external validity will be futile, as there will be no internally valid inference to be generalised in the first place. Striking a reasonable balance between internal and external validity of crowd experiments should be made a major design consideration.

# 1. Introduction

The importance of experiments in the field of crowd dynamics is being exceedingly recognised more than ever before. This has been documented by the unprecedented number of experimental studies that have been published over the last few years, outsizing, by far margin, the entire body of experimental work that had been undertaken over many more years prior (Haghani, 2020a, b, 2021). The experimental approach is being treated as a step forward relative to pure computer simulation. While this approach is often being regarded as a necessity for validating outcomes of numerical models or finetuning their predictions, they are not exempted from the question of validity.

Within the field of crowd dynamics, most of us see experimentation as a benchmark for validity of our models and theories. While we often take pride in taking those theories into the laboratory and testing them against experimental observations, it is not an unfair inquiry to question how valid these experiments themselves are. This question, in my personal experience, is more frequently raised by scholars from outside the crowd research domain, as compared to the researchers who identify with this field. We may have noted such concerns being expressed to us when presenting experimental outcomes to a diverse audience or when receiving review feedback from our peers who may not necessarily be familiar with such experiments. Typical questions raised may include: "it is not clear to me how valid such data is from a simulated evacuation experiment, this is not a real-world emergency" or "have you considered validating your data" or "why do you do experiments, why don't you use CCTV footage of real evacuations, there must be plenty out there". Another frequent comment that I receive concerns the use of university students as subjects of such experiments. This is put to me in statements such as "how much do you think your results will be limited by the fact that you didn't have elderly people or people of different cultural backgrounds in your crowd", a question which, as we discuss later, relates directly to only one of several dimensions of validity. And I believe that many in the experimental sector of this research field might have had similar experiences from the audience of their research. While I have not been able to document such verbal questions verbatim, despite facing them almost every time after presenting experimental findings, there are direct quotes that I have documented from the assessors of previous research grant proposals. Here, I provide three examples of such statements from three separate assessors:

> Quote #1: "*My only concern is that the experiment design may not be good enough to capture the human factors*"
>
> Quote #2: "*Whether this will go beyond theory and translate into practical benefits for crowd evacuation scenarios in emergencies is unclear. Obviously, all of crowd research suffers from the disconnect between controlled experiments and emergency scenarios: will the experimentally derived insights apply in real-world scenarios?*"
>
> Quote #3: "*The uncertainties lie in whether the knowledge developed under controlled laboratory conditions can be effectively applied in real cities where behaviour is mediated by so many additional factors.*"

Many crowd researchers, particularly if they have engaged in publishing experimental work, often see experimental research as a reasonable benchmark for their scientific inquiries. However, this view is not shared universally. This is rather striking, in that, psychologists and economists frequently use experimental data in their investigations. Economists are often divided on the suitability of creating markets in the laboratory and experimenting financial decision-making in artificial settings rather than real-world markets. Nevertheless, the mere existence of *experimental economics* as a lively and established field is a testament to the fact that experimentation is a relevant and an acceptable form of scientific enquiry, at least among a substantial body of economists. Another sign of such recognition is the fact that, in many instances during the recent years, *experimental/behavioural* economists have been recipients of Nobel Prizes. For psychologists, often laboratory experiments are the only available tool for addressing their research questions, and as such, experiments are long-established and an accepted approach among psychologists. The relative acceptability of experimentation is not limited to mother fields such as psychology or economics, as there are applied and derived fields of research that have been employing such approaches for a considerable period of time. The ubiquitousness of driving simulator experiments in the domain of road safety in transportation research, for example, is only one of such cases. Clearly, a drive in a simulator is not equivalent to a drive in the real world and there are fundamental ecological differences between the two (hence, the names 'simulation' and 'experiment'). From that perspective, such experiments do share undeniable common elements with the type of experiments that crowd researchers undertake. One group creates conditions of driving under the influence of alcohol and the other may create conditions of a stressful building evacuation under a simulated emergency in a virtual-reality setting. But why is it that a simulated driving experiment is often deemed a more acceptable form of experimentation, whereas a simulated evacuation is sometimes viewed with scepticism with respect to its validity? Or more generally, why is it more acceptable to create artificial markets or social settings in laboratories—sometimes in gamified interactive situations, sometimes in novel unexperienced situations, sometimes in situations involving stress and emotions— in order to observe human behaviour, perception and decision-making, but not so acceptable when one creates an evacuation or crowd decision-making situation in comparable settings? What makes immolation of crowds and evacuations more outlandish than these other forms of experimentation?

Compared to psychology experiments or even those of driving simulation, experimental work in crowd dynamics is relatively young ([Haghani, 2021](#)). We have only seen continuous streams of experimental papers in this field over the last 10-15 years. To some degree, this may explain why crowd experiments are subject to more scepticism with respect to their validity. More people have observed and been exposed to driving simulator experiments, and road safety policies have been made based on such outcomes. This may per se create a certain level of acceptability. Another possible reason is that, being around as a method for scientific inquiries for a much longer time, driving simulator experiments may have been more frequently tested against counterpart experiments or field data. It could also be that more established and more rigorous protocols for conducting such experiments and assessing their validity have been developed over the years. Or perhaps, all of these are aggravated by the fact that many of us in the field of crowd dynamics are from physics, engineering or computer science backgrounds. Perhaps in the eyes of onlookers, we can just not be trusted with designing and executing good behavioural experiments.

Nevertheless, here we argue that, while crowd experiments are no more outlandish than experiments of psychologists and behavioural economists, and while in fact there are striking parallels between them, they are still subject to the question of validity. Given the unique features of experiments in crowd dynamics, the variety of their purposes/topics, the variety of their methodologies and the variety of their design/analysis approaches, it is believed that the notion of validity deserves to be considered with more nuance within this specific field of research. This would be essential for multiple reasons. Such efforts will make it more readily possible to assess validity of experiments, on a case-by-case basis, using a common language and standard set of criteria. This will take us closer to having more established protocols in experiment design in this field. It would also allow researchers of this field to consider the question of validity during the design of their experiments and recognise particular dimensions of validity that are more important to the purpose of their experiments. But most importantly, it would enable us to better articulate how valid and generalisable are experiments are, how their results should be interpreted and how far such results could be extrapolated and stretched.

The current work could be considered as a first step towards this goal. Considering the existing theories regarding the notion of validity in a general domain of social, behavioural and medical sciences, here, we decompose various dimensions of validity as they pertain to crowd experiments. The goal is to recognise the multitude— and sometimes, competing—aspects of validity that are relevant to each crowd experiment and to establish how their significance may vary depending on the primary purpose of the investigation. Discussions centre around these questions: How can one consciously create a justifiable balance between these competing elements of validity during the design? How can potential factors that may jeopardise validity of crowd experiments be identified and mitigated during the design? Is field data inherently superior to experimental data when it comes to the matter of validity?

## 2. Empirical versus numerical research in crowd dynamics

Traditionally, pedestrian dynamics researchers employ two general avenues of research for their investigations, one that is predominantly based on the use or development of numerical simulation models, i.e., *numerical methods*, and one that is predominantly based on data collection, i.e., *empirical methods*. The former embodies what we often refer to as methodological studies, a class of studies which used to be rather disconnected from the empirical stream. The emergence or adoption of new methodologies such as coarse network-based models (see O'Leary and Gratz (1981), Turner (1984) and Kendik (1983) for examples of pioneering work in this category), or macroscopic fluid-type models (see Henderson (1974), Thompson and Marchant (1995) and Henderson (1971) for examples of pioneering work in this category), and subsequently agent-based microscopic models (see Helbing and Molnar (1995) and Burstedde et al. (2001) for examples of pioneering work in this category) as well as various methodologies of numerical optimisation of evacuations are all examples of this line of work. From this perspective these numerical investigations can be categorised as *descriptive* and *prescriptive* (also known as, *optimisation*), a differentiation that was highlighted with more depth in an earlier work (Haghani, 2020c). On the other hand, for as long as crowd dynamics existed, researchers and particularly human-behaviour-in-fire (HBiF) researchers who were pioneers in this category (see Haghani (2021) for historical perspectives) have employed observational/empirical methods for their investigations. A pioneering example of this line of investigation can be found in the work of Kinateder and Warren (2021). In earlier work, I made distinctions between the cohort of empirical studies whose observations are obtained from naturally occurring environments from of experimental interventions, i.e., *field methods* Haghani (2020b), and those where the investigator create some artificial settings, recruit participants and exerts certain levels of intervention, i.e., *experimental methods* Haghani (2020a).

Up until 10 years ago, the literature of crowd dynamics used to display a disconnect between empirical and numerical work. More recently, the border appears to be fading away, with researchers nowadays employing a combination of methods for their investigation. Due to the recognition that empirical work has received during the recent years, most modern numerical studies can only be published if they demonstrate some form of connection to real-world or empirical settings. Purely numerical papers in this field, especially those of descriptive nature, are becoming increasingly rare. Nowadays, authors of numerical papers are often expected to establish some form of connection to the empirical front, e.g., present some level validity investigation for their models. Similarly, in many empirical papers, we observe that researchers have employed the empirical insight to develop or modify a numerical methodology or test a previously established methodology. As a result of this paradigm shift in crowd research (highlighted in an earlier work (Haghani, 2021)), the distinction between numerical and empirical research in crowd dynamics is no longer necessarily absolute. For contemporary papers, this distinction can often only be made by determining the component that makes up the dominant aspect and main focus of the study.

Out of the variety of studies in crowd dynamics that we recognised above, here, the focus is only on one particular class, experimental studies, studies where the experimental component is the dominant focus. Empirical crowd research whose underlying data is sourced from natural settings are not of concern here. Our discussions particularly focus on the notion of validity in experimental papers of pedestrian, crowd and evacuation dynamics research. While observations obtained from the CCTV footage of crowd reaction in a mass shooting or pedestrian trajectory data obtained from an actual road intersection (free of any experimental intervention) or retrospective interview studies with survivors of mass emergencies (Drury et al., 2016; Drury et al., 2009a, b) can all be questioned on account of data accuracy, they are not the subject of our discussion here (as they are deemed field data). Here, we only focus on providing a clearer description of the notion of validity in relation to experimental crowd research. The aim is to better recognise dimensions of validity, the trade-off that often exists between these dimensions in the design phase of crowd experiments, and to determine the basis on which these trade-offs should be made. We discuss why it is essential that the researcher is conscious of such trade-offs and makes prioritisation. Our discussions also provide insight into how claims of individual experimental studies should be assessed in relation to these validity dimensions, and how threats to various components of validity can be recognised during the design and execution phase (and mitigated).

## 3. Categorisation of crowd dynamics experiments based on method and purpose

In previous work, experiments of crowd dynamics have been categorised based on the methodology of experiments. This namely includes studies that (a) employ animal or insect groups as entities of experiments, (b) studies conducted in virtual or augmented reality settings or using hypothetical choice surveys, (c) studies labelled as lab-in-the-field (or field-type) experiments where groups of human participants perform experimental tasks in non-virtual but artificial settings, and (d) evacuation drill experiments where groups of human participants perform the tasks in real built environments. While the difference between the third and fourth category could, in some cases, be subtle, they can often be distinguished by the fact that evacuation drill experiments have minimal experimental intervention and often do not have a large number of repetitions and/or alternation of design factors (as we observe in most field-type experiments). However, here, we argue that this categorisation has not much bearing on the issue of validity. As we discuss later, any of these four types of experiments can potentially score high on various dimensions of validity when properly designed and executed. We argue that what matters the most in relation to the assessment of validity in crowd dynamics is the *purpose* of experiments.

We argue that experiments of crowd dynamics could each be attributed to one of the following four categories: (i) those whose main aim is to test a specific behavioural theory or hypothesis, (ii) those that are designed for exploring empirical behavioural regularities (with no pre-defined hypothesis), (iii) experiments where certain behaviour is instructed (rather than observed) with the aim of determining the impact of instructed behaviour on system-level measures, and (iv) experiments designed specifically for calibration of numerical models.

In simple terms, in the first two categories, our investigations concern the behaviour of individuals, as we aim to learn or often model their behaviour. Whereas, in the third category, we have specific types of behavioural (and/or architectural) intervention in mind and wish to know how the system is impacted as a result of that intervention. Note that this third class of experiments could be hypothesis-based too (and in fact, many of them are). Yet, they are often clearly distinguishable from the first two classes on the account that the entity of interest in this class of experiments is not individual behaviour. The focus is rather on system-level metrics. Also, note that in both Type (i) and (ii) experiments, the investigator may incorporate different *treatments* in the design, where modifications are made to the behaviour of people or the architecture of experimental environment. However, so long as the main purpose of the design remains on learning individual's behaviour (rather than measuring system performance metrics), then the experiment cannot be confused with Type (iii). Furthermore, in relation to the third category, it is not always the behaviour that is instructed. Sometimes the modification concerns architectural elements of environment in which the experiment is conducted. But again, so long as the design aims to quantify the impact of that modification (be it behavioural or architectural) at the system (or macro-scale) level, then the experiment could be distinctly categorised as Type (iii).

In relation to Type (iv) experiments, it might be argued that any attempt in calibrating a model is to some degrees an attempt to find behavioural regularities, and from that perspective, Type (ii) and (iv) may not be necessarily differentiable. The main criterion of differentiation here would be that when an experiment is designed for model calibration, there is often a pre-specified structure for model formulation and the purpose is just to gather enough data to feed the model and obtain estimates of the parameters. Whereas in Type (ii) experiments (i.e., exploratory experiments), the analyst may not necessarily have a pre-existing theory-based model in mind. A statistical model might be developed as part of the investigation but that itself is part of the exploration. Finally, it is important to distinguish four further sub-categories of Type (iv) experiments: those whose aim is comprehensive/global calibration of a pre-existing model (Type (iv-a)) and those whose aim is calibrating only one or a certain subset of parameters of a pre-existing model, i.e., a partial or local calibration, (Type (iv-b)). The third sub-class of this category are experiments designed for *directly measuring* a quantity rather than calibrating a parameter (Type (iv-c)). These experiments (i.e, Type (iv-c)) are rare in the literature of crowd dynamics, but examples of them do exist, as we will demonstrate in subsequent sections. A fourth sub-class of this category can be identifiable where the experiment is not meant to produce estimation of any parameters, rather is purely used to assess model accuracy and performance (Type (iv-d)).

To elucidate this issue, here, examples of crowd dynamics experiment of each kind have been sampled from the literature. The studies listed in Table 1 will also be mentioned in subsequent sections as references when discussing various dimensions of validity.

**Table 1** Examples of different categories of crowd experiments based on their purpose of investigation.

| reference | title | method | category |
|---|---|---|---|
| Tong and Bode (2021) | The value pedestrians attribute to environmental information diminishes in route choice sequences | virtual reality | (i) |
| Haghani and Sarvi (2019a) | 'Herding' in direction choice-making during collective escape of crowds: How likely is it and what moderates it? | lab in the field | (i) |
| Kinateder et al. (2014) | Social influence on route choice in a virtual reality tunnel fire | virtual reality | (i) |
| Bode et al. (2015a) | Disentangling the impact of social groups on response times and movement dynamics in evacuations | lab in the field | (i) |
| Haghani and Sarvi (2017b) | Stated and revealed exit choices of pedestrian crowd evacuees | lab in the field | (ii) |
| Liao et al. (2017) | Route choice in pedestrians: determinants for initial choices and revising decisions | lab in the field | (ii) |
| Moussaïd et al. (2012) | Traffic instabilities in self-organized pedestrian crowds | lab in the field | (ii) |
| Lovreglio et al. (2014) | A discrete choice model based on random utilities for exit choice in emergency evacuations | virtual reality | (ii) |
| Garcimartín et al. (2016) | Flow of pedestrians through narrow doors with different competitiveness | lab in the field | (iii) |
| Zuriguel et al. (2020) | Contact forces and dynamics of pedestrians evacuating a room: The column effect | lab in the field | (iii) |
| Sieben et al. (2017) | Collective phenomena in crowds—Where pedestrian dynamics need social psychology | lab in the field | (iii) |
| Shahhoseini and Sarvi (2019) | Pedestrian crowd flows in shared spaces: Investigating the impact of geometry based on micro and macro scale measures | lab in the field | (iii) |
| Dias and Lovreglio (2018) | Calibrating cellular automaton models for pedestrians walking through corners | lab in the field | (iv-a) |
| Lovreglio et al. (2015) | Calibrating floor field cellular automaton models for pedestrian dynamics by using likelihood function optimization | virtual reality | (iv-a) |
| Guo et al. (2020) | Characteristics of pedestrian flow based on an improved least-effort model considering body rotation | lab in the field | (iv-b) |
| Yamamoto et al. (2019) | Body-rotation behaviour of pedestrians for collision avoidance in passing and cross flow | lab in the field | (iv-b) |
| Zhao et al. (2019) | Quantitative measurement of social repulsive force in pedestrian movements based on physiological responses | lab in the field | (iv-c) |
| Xiao et al. (2019) | Investigation of pedestrian dynamics in circle antipode experiments: Analysis and model evaluation with macroscopic indexes | lab in the field | (iv-d) |

## 4. Dimensions of validity and their interpretations in crowd dynamics

The notion of *validity* (Cook and Campbell, 1979) is a very broad term and, in general, refers to the extent to which results of as study translate to the true state of some observed phenomenon, e.g., the extent to which results of a simulated evacuation experiment translate to the actual behaviour of humans in an evacuation scenario in real-world emergencies. This terminology is commonly shared across most domains of social, physical and medical sciences (Bishop and Boyle, 2019; Karren and Barringer, 2002; Mariel et al., 2021; Parady et al., 2021; Slack and Draugalis Jr, 2001; Toubia et al., 2003; Vossler et al., 2012). However, in some contexts of medical or social sciences, the notion of validity could be used in relation to a *measurement instrument* or *scale* (Kimberlin and Winterstein, 2008) rather than an *experiment*. Here, we refer to this notion only in relation to behavioural experiments (Khan, 2011; Lynch Jr, 1982; Winer, 1999).

Across social sciences, in particular, various models of validity have been proposed while the model of *internal* and *external* validity still remains the most popular (Haghani et al., 2021). Attempts for relabelling these dimensions or proposing alternative dimensions have been made in previous work such as that of Campbell (1986) entitled "Relabelling internal and external validity for applied social scientists" where alternative labels such as 'local molar causal validity' and 'principle of proximal similarity' were proposed. Here in this work, we adhere to the established model of *internal* and *external* validity and its interpretation in relation to crowd experiments. Within this realm, *internal validity* of a crowd experiment refers to the extent to which the experiment can reveal genuine causal relations (or the extent to which it can reveal the absence of causal relationship). Maximising this feature of a crowd experiment could often necessitate abstraction, simplification and exertion of high level of experimental control. *External validity* of a crowd experiment, on the other hand, concerns the extent of generalisability of findings to the contexts beyond the experiment setting. Often, these two features of a crowd experiment could conflict and create a trade-off. This conflict and its implications on experiment design as well as interpretation and assessment of experimental findings in crowd dynamics essentially makes up the main motivation of this work. Enhancing external validity of experiments often can only come at the cost of compromising internal validity and losing experimental control, and vice versa. In other words, in vast majority of cases, the investigator cannot maintain high levels of both dimensions at the same time. The more the investigator controls for extraneous factors, the more he/she relinquishes the ability to generalise the findings to a broader context. Therefore, it would be essential to consciously recognise the dimension of validity (i.e., internal or external) that matters the most in relation to each investigation and tailor the design of the experiment accordingly.

## 5. Internal validity in crowd dynamics experiments

### 5.1. Experiment type and its relation to internal validity

Maximising internal validity of an experiment means that the investigator establishes confidence that he/she is measuring a genuine cause-and-effect relationship and ensuring that the observed relationship is not confounded with other factors. This often requires isolating the variables/factors of interest from all other potential contributing factors, and therefore, it may entail simplification and design artificiality. When it comes to crowd experiments (where features of the build environment within which pedestrian interactions take place is a major factor), such simplification often translates to artificiality of the environment and compromising experimental fidelity. Gaining internal validity at the expense of compromising generalisability in crowd experiments, e.g., experiments in highly simplified settings rather than generic and natural settings, is most justifiable when we either have a specific theory or hypothesis in mind to test (experiments Type (i)) or have a specific subset of parameter(s) of a model to calibrate (experiments Type (iv-b)), or have specific quantities to measure (experiments Type (iv-c)). It is desirable to maintain contextual and environmental fidelity too, but a certain level of compromise often becomes inevitable in order for the investigator to be able to isolate the factor/metric/variable of interest and test the hypothesis/theory, or measure the intended variable. The experiments shown in Figure 1 are all examples of such cases. The experiment of Tong and Bode (2021) (Figure 1 (1)), for example, was primarily designed to test the hypothesis that "as pedestrians make more route choices in sequence, they attend less to posted information". This is a very specific theory or hypothesis, and therefore, requires an experimental environment in which the effect of all other potential contributing factors been taken out of the picture or at least been minimised. In terms of the causal relationship, here, the cause is repetition in route choice making and the effect is lower attention to information. Such design need not be in a virtual-reality setting. But regardless of whether such experiment is conducted in the virtual-reality or field-tyle setting, testing the said hypothesis via an internally valid experiment would make certain levels of simplifications inevitable.

In the example of Tong and Bode (2021), establishing internal validity essentially means that the investigators had to be able to demonstrate that pedestrians attending less to direction information was indeed a result of making more and more route choices and not any other factor. Whether (or more precisely, the extent to which) we can claim that this observation is generalisable to pedestrian route choice in real-world settings determines the external validity of this experiment.

Similarly, in Haghani and Sarvi (2019a), the investigators had a specific hypothesis in mind and that was the so-called 'herd theory' stating that as perceived level of urgency (the cause) increases in an evacuation, people tend to follow the decision of majority (the effect). Testing this hypothesis required creation of an experimental setting in which, other than urgency level and social influence, the impact of every other factor is minimised. The very artificial and simplified setting in Figure 1 (2), which does not bear much resemblance to any actual building. In Bode et al. (2015a) (Figure 1 (3)), an underlying hypothesis of the design is that pedestrians in social groups show a delayed reaction to an emergency evacuation signal before they initiate their movement. Here, the cause is being in a social group and the effect is delayed response.

We should also draw a contrast between the experiment of Tong and Bode (2021) and that of Haghani and Sarvi (2019a). Both are designed to test a theory or assumption. In Tong and Bode (2021) this is an assumption that authors made based on their intuition and perhaps previous observations and possibly gathering cues from the literature. Their experimental observation indicated that their assumption cannot be ruled out, hence, taking us one step closer to believing that pedestrians stop attending to signs and information as they make more route choices. In other words, the authors theorised an assumption, then offered experimental findings in support of that assumption. In Haghani and Sarvi (2019a), however, the investigators were focused on an established assumption that had propagated heavily in the literature for years (without being empirically tested). The assumption was that higher levels of perceived urgency motivates people to follow the crowd.

In a previous experiment, (Haghani and Sarvi, 2017a) observed indications that the herd behaviour theory may not universally hold true. This provided the motive for designing a secondary experiment, to specifically test this assumption under higher degrees of internal validity compared to that of Haghani and Sarvi (2017a). This experiment again produced findings against the herd behaviour theory. This is an example of a case where an untested theory is essentially challenged by experimental observations In other words, while Tong and Bode (2021) and Haghani and Sarvi (2019a) are both hypothesis-based experiments, one demonstrated *existence* of a causal effect and the other demonstrated *absence* of a relation. The reason these two examples were contrasted is to highlight the fact that while behavioural experiments are good for disproving a theory and ruling out their universality, they are not designed to *prove* theories. A single (internally valid) experiment can disprove a theory or its universality. When it comes to proving a theory, however, we need accumulation of evidence. Every experiment that is conducted and is unable to reject a theory makes us more confident that the theory holds true. In other words, it takes one (internally valid) experiment to rule out universality of a theory, but single experiments are generally not meant for proving any theory. Accumulation of empirical evidence in favour of a theory/assumption can generate more confidence in the theory. This may be important when interpreting positive and negative findings in hypothesis-based crowd experiments.

Another form of empirical enquiry in which internal validity plays an important role and requires simplified experiments is when we seek to calibrate a certain parameter of a pre-existing model or measure a specific quantity within a model. Examples of these can be seen in the work of Guo et al. (2020) (Figure 1 (4)) where specific parameters related to body rotation of pedestrians within a general framework of a *least-effort model* were to be calibrated. Similar concept can be found in the work of Yamamoto et al. (2019) (Figure 1 (6)). As can be seen, the design of both experiments aims to minimise the effect of a range potential factors in order for the investigator to be able to effectively isolate and study the factor of interest (here, the body rotation parameters). In both cases, not much social influence (crowding) is presented. This means external validity is compromised as it limits the ability of the investigator to generalise the results to settings in which crowding and social influence are at play. Another example is the experiment reported by Zhao et al. (2019) where the aim was to measure *social repulsive force* (within the realm of the *social force* pedestrian model) and that required designing a highly artificial setting that may not have much resemblance to real-world pedestrian environments in order to isolate the variable of interest that needs to be measured.

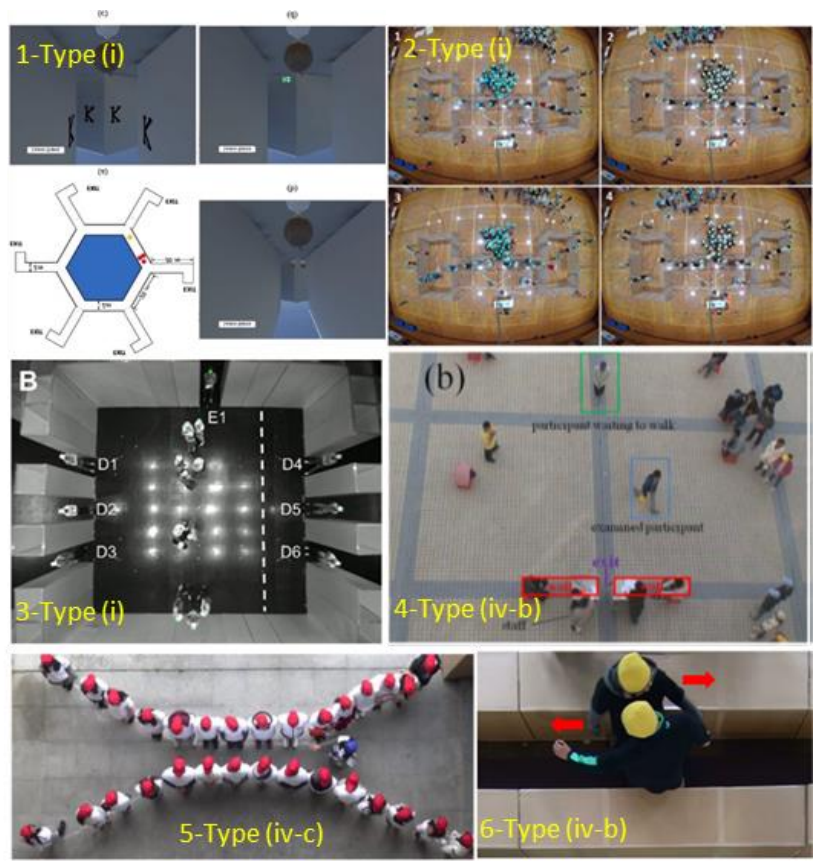

**Figure 1** Examples of crowd experiments where gaining internal validity required simple and artificial designs: (1) Tong and Bode (2021), (2) Haghani and Sarvi (2019a), (3) Bode et al. (2015a), (4) Guo et al. (2020), (5) Zhao et al. (2019), (6) Yamamoto et al. (2019).

It should be noted that the above discussion was not meant to convey the message that Type (i) (or Type (iv-b/c)) experiments need to be conducted in overly simplified settings. The suggestion is that lack of experimental fidelity (or what we refer to, in the following section, as *ecological validity*, i.e., the extent to which situation of a crowd experiment reflects real life) could be more acceptable or warranted in relation to this type of experiment. However, if internal validity can still be achieved at the same time as maintaining ecological validity, then oversimplification should be avoided. There areexamples of hypothesis-based experiments in crowd dynamics where experimental treatment is fairly realistic and not oversimplified. Take experiments of Kinateder and Warren (2021) (Figure 2 (1)) or Kinateder et al. (2014) (Figure 2 (2)) which are both hypothesis based. Both experiments were conducted in virtual-reality settings, and as far as ecological validity goes within the realm of virtual-reality experiments, both are fairly realistic and not overly simplified. That said, the decision on the level of simplification when dealing with hypothesis-based experiments, is a call that the investigator has to make. In such cases, it would be ideal to keep the compromise of ecological validity to a minimum extent and find a reasonable balance between controllability and realism in the design.

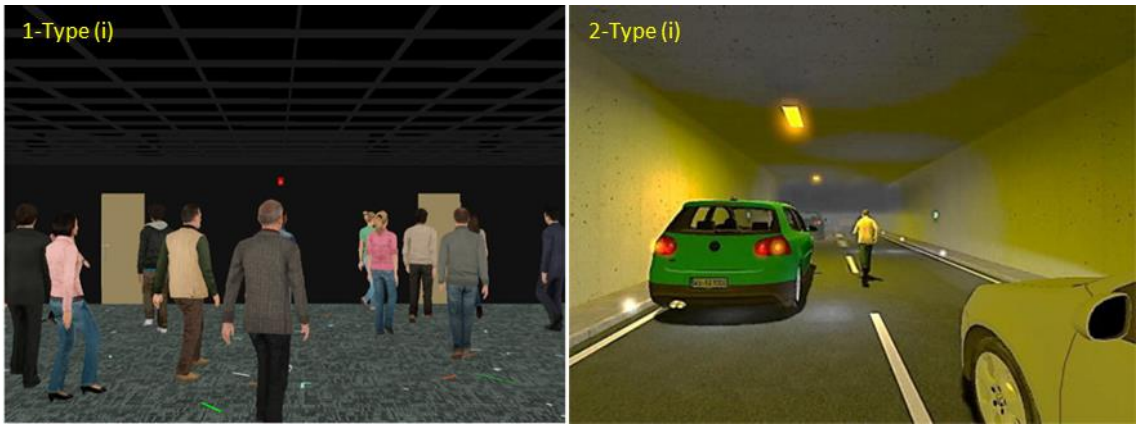

**Figure 2** Examples of hypothesis-based crowd experiments where the environmental design is fairly realistic. (1) Kinateder and Warren (2021), (2) Kinateder et al. (2014)

On the opposite side, if an experiment is not designed for hypothesis or theory testing or calibration of specific parameters within a given model, then a highly artificial design would be hard to justify and defend. For example, in Haghani and Sarvi (2019a), if the experiment was meant to investigate exit choice behaviour of pedestrians during evacuations, in general, then the design (Figure 1 (2)) would not be easily justifiable, as the experimental setup lacks enough genericness (or naturality) for such purpose. There are many potential factors that influence exit choice but not at play in such simple and symmetric (i.e., highly artificial) layout. Those effects have been deliberately taken out of the picture (e.g., by creating highly symmetrical exit choice situations where the factor of spatial distance has no role to play). Similarly, take the experiment reported in Xiao et al. (2019) and Xiao et al. (2021). The experiment creates a circular antipode setup, with the aim of investigating pedestrian conflict resolution behaviour. Such setup is very specific and not natural. Pedestrians barely ever face such symmetric and circular setting in real life when walking on crowded spaces. And at the same time, the experiment is not driven by specific hypotheses and is, as a result, mostly exploratory. The circular setup and a range of its variation was created (Figure 3) and then patterns were discovered by analysing individuals' behaviour, hence the suggestion that the experiment is largely exploratory. A secondary purpose of the experiment is also model validation for which ecological validity is also a desirable (or required) attribute. This could be an example where the compromise of external validity (more specifically, ecological validity) in favour of gaining internal validity could be hard to justify.

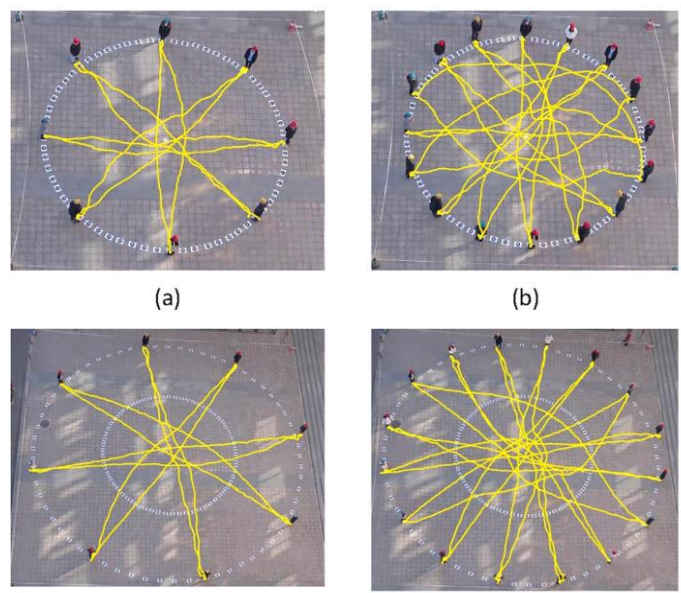

**Figure 3** Examples of an exploratory (type (ii)) experiment in a specific and artificial setting, Xiao et al. (2019).

*5.2. Dimensions of internal validity in crowd dynamics experiments*

Two dimensions for internal validity has been recognised. One is *content validity* which, in the context of crowd experiments, could refer to good practice in design and execution. This will be further elaborated when we enumerate threats to internal validity. Essentially, any effort to identify and minimise potential threats to internal validity of crowd experiments could be considered an element of good practice. Therefore, experiments where these precautions have been taken adequately would naturally score high on this particular dimension of internal validity. The other dimension is *construct validity* which, in the context of crowd experiments, can be best described as to whether instructions of experiments were adequately followed by participants. This concept generally means whether the experiment measures what it claims to be measuring. Since our external measurements in crowd experiments are often fairly accurate, if we make sure that participants adequately followed the instructions, then we can claim that we are measuring what we intend to measure.

For example, in Haghani and Sarvi (2019a), the investigation was looking at the impact of perceived urgency on the tendency to follow the direction chosen by the majority. Therefore, a theoretical construct that was to be created in the experiment, and that was the level of simulated urgency. This was created through proxies and by introducing elements such as auditory stimulus associated with urgencies and creating incentive-compatible competitions between participants, assuming that they create an elevated level of perceived urgency (compared to the baseline treatment where these elements were absent). To the extent that those stimuli successfully created the intended level of elevated urgency (compared to baseline), the experiment can be claimed to possess construct validity[1]. This particular dimension of internal validity, construct validity, could in general be most relevant to Type (iii) experiments where the creation of the construct of interest depends on whether instructed behaviour was followed by participants. For example, a faster-is-slower experiment is an example of a Type (iii) experiment (e.g., see the experiment of Garcimartín et al. (2016), Figure 4 (1)). The creation of the construct of interest in this experiment relies on whether participants applied the high and low level of vigour and physical force when exiting, and that is something that can be measured though proxies such as density maps. Experiments reported in Sieben et al. (2017) (Figure 4 (2)) or Haghani et al. (2019) (Figure 4 (3)) are also of this type for which the matter of internal validity relies mostly on whether instructions were followed by participants and whether the construct of interest were successfully created as a result. Sieben et al. (2017), for example, instructed participants to imagine they are at the entrance of a concert venue by their favourite band where only the first 100 people were able to access the concert. Any subsequent behaviour and casual effect (here, the effect of architectural arrangement at the entrance on people's behaviour) relies on whether the initial instruction was followed. Had the participants not taken these instructions seriously at all (a rare occurrence in experiments, as participants almost always try to be "good participants"), then the change in the type of architectural design (i.e., semicircle versus corroder) would have probably shown no effect, suggesting (falsely) inexistence of a causal relationship. This would have essentially nullified the internal validity of the experiment. Another threat would have been participants applying the instructions differently across the two architectural treatments (e.g., learning from the first round (baseline) and modifying their behaviour in the second round of the trials (treatment)). That would have also jeopardised internal validity of the inferences. In Haghani et al. (2019), the creation of the theoretical construct (i.e., the faster-is-slower effect under non-aggressive levels of pushing) relied on participants being able to demonstrate three types of behaviour (i.e., three treatments). One in which they were instructed to assume they were existing a boring lecture while not being in rush, one where they assumed they were entering a train and trying to secure an empty seat, and one where they assumed they were escaping a danger. In all treatments, the instructions disallowed aggressive pushing. The extent to which participants could follow these instructions and maintain them across the repetitions determines the construct validity of this experiment. That is why it was essential that the instructions be repeated to participants at every round of the trials.

---

[1] Had this experiment been conducted in a virtual-reality environment, it could have been possible to measure physiological response and further indicators of stress and ascertain whether the theoretical construct was successfully achieved by the design of the experiment. This is another example of higher degrees of controllability that can be achieved when performing experiments in virtual-reality environments as opposed to field-type settings. In the absence of such physiological measurements, less representative measures such as the movement speed of people were used to determine whether the sense of urgency was created by the experimental treatment.

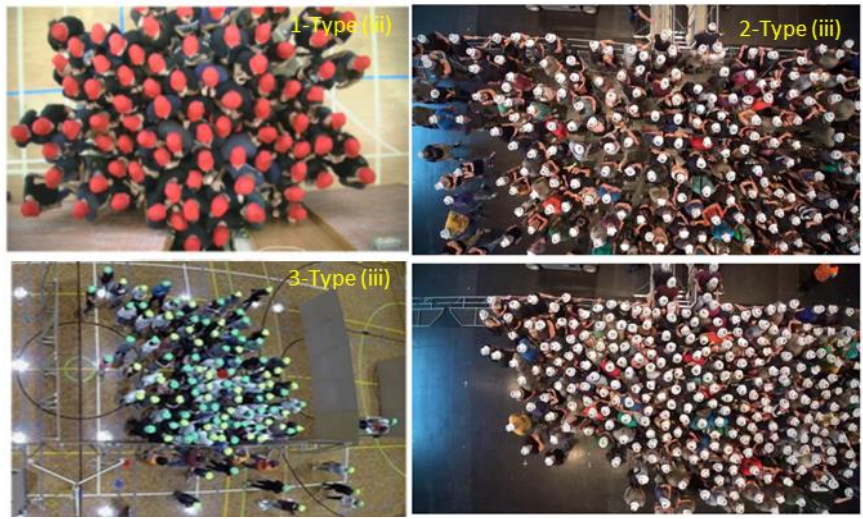

**Figure 4** Examples of crowd experiments where construct validity is the primary determinant of internal validity. (1) Garcimartín et al. (2016), (2) Sieben et al. (2017), (3) Haghani et al. (2019).

*5.3. Threats to internal validity of crowd dynamics experiments*

A variety of factors, depending on the context and type of a crowd experiment, can potentially jeopardise internal validity of crowd experiments. In general, anything that reduces study's ability ability to establish a relationship between the independent and dependant variables reduces internal validity. For example, consider (a hypothetical) between-group design that aims to test the faster-is-slower assumption, where one group performs the high motivation task and the other performs the low motivation task. If there are significant differences between these two groups in terms of the attributes that are relevant to discharge rate from an exit (e.g., one group is composed of adults with significantly larger body sizes), then the experiment will suffer from *selection bias*, a threat to internal validity.

Consider a crowd bidirectional flow experiment conducted shortly after restrictions related to Covid-19 are lifted in a country, while people are still mentally primed by the notion of physical distancing and unintentionally (without the experimenter instructing them to do so), maintain larger (than usual) physical spaces to other people during the experiment. This is an event unrelated to the experiment that can influence its outcomes. In this case, the experiment suffers from the threat of *history* to internal validity.

Consider a hypothetical crowd experiment in a maze type environment where the aim is to investigate whether placing signs can reduce evacuation time. We perform 10 repetitions without posted signs and then perform another 10 repetitions under the sign treatment. During the first treatment however, participants might have become more familiar with the maze, and any potential reduction in evacuation time in the second phase could partly be the result of this higher level of familiarity. Such experiment suffers from *testing* effect another threat that can create an unintended confound thereby jeopardising internal validity.

Consider a faster-is-slower experiment. We conduct 5 repetitions under the slow condition followed by 5 other repetitions under the fast (i.e., high motivation) condition, with a fixed doorway width. But we are using a temporary structure to create the door opening of the desired width. However, in transition between the two phases (i.e., slow to fast), the physical setup incurs unintended changes for some reason making the doorway a few centimetres wider before the start of the second 5 repetitions. This experiment will suffer from the threat of *instrumentation* to internal validity. An alternative example could be a field-type experiment (say, designed for investigating fundamental diagram of pedestrian movement) where we have multiple repetitions, and halfway through the experiment, the camera's calibration setting is disturbed.

Consider an experiment where two groups of participants are performing similar evacuation tasks in parallel. The only difference is that we have trained one group to perform their tasks in a certain way (i.e., we have modified the behaviour of the treatment group through instructions). The group sizes are at the beginning the same and the aim is to compare their overall evacuation time across the control and treatment groups. If some participants drop out of the experiments from one group after a few rounds, this will affect our measurements of total evacuation time and affects the accuracy of comparisons between the two groups. Such experiment would suffer from the threat of *attrition* or *morality* to its internal validity.

Consider an evacuation experiment. We have a crowd of 100 participants, and they perform an evacuation task five times. Then, we separate a subset of 50 participants and instruct them to modify a certain aspect of their behaviour. In other words, we train those fifty people. The aim is to understand how much more efficient the system would be if 50% of the crowd adopted the "desired" behaviour. These trained participants go back to their group and mix up with them, before the start of the next rounds. But in the meantime, they also have an opportunity to communicate with their fellow (naïve) participants. They might share what instructions they received, and this may prompt some of the members of the other group to also modify that behaviour. Such experiment would suffer from the threat of *social interaction* to its internal validity.

Assume that a series of evacuation experiments are designed to investigate the effect of an architectural modification on evacuation time. We perform 10 repetitions under the first architectural treatment when participants are fresh and full of energy. But it takes us 1 hour to prepare the next setup and, during this time, participants get hungry, bored and tired, and as a result, they perform the tasks of the next treatment with less vigour (compared to the first round). This confounds with our evacuation time measurements and any subsequent comparisons across treatments. This constitutes a threat of *maturation* to internal validity, the possibility of unintended mental/physical changes occurring within participants, thereby, creating a confound.

*5.4. The choice of experimental method in crowd dynamics and its relation to validity*

In this section, our examples of crowd experiments remained confined to virtual-reality experiments and field-type experiments. A point that needs to be considered is that, while virtual-reality settings often provide more controllability to the experimenter, it is not a universal rule that they are ecologically less realistic than field-type experiments. Although it is true that field-type experiments take place in an actual physical environment and any interaction with other pedestrians is real (as opposed to interactions with virtual humans/avatars), a field-type setting could be oversimplified, and a virtual-reality setting could be made ecologically realistic. Therefore, there should does not exist any absolute and universal ranking between ecological realism of these two methods. Each could be made realistic or highly simplistic. The choice between the two also involves a range of other factors in addition to the issue of ecological validity. The underlying question of the experiment itself could be a major factor. Are we dealing with a problem where physical interaction and contact between people are critical? Then in that case, the more suitable choice would be taking the lab to the field. Also, these two types of experiments may pose different levels of logistical challenges and financial costs to the experimenter ([Haghani and Sarvi, 2018](#)). Another matter could be convenience, availability of facilities, expertise, previous experience of the investigators in employing these methods etc. Therefore, neither has an inherent superiority in experimental crowd research. Similar point can be made about evacuation drill experiments and their relation to virtual-reality and field-type experiments. None of them score inherently higher on the matter of validity and the choice of the method should be made in consideration of a range of factors, including the question of interest.

The fourth method of crowd experiments, however, i.e., experiments using ants or animal groups were never received a mention in our previous examples. As far as the matter of validity is concerned, such experiments can still be *internally* valid. Although the component of *construct* validity would not be easy to maintain as one cannot be sure that the animals exactly followed the "instructions". However, so long as the investigator is not seeking generalisation and limits interpretations of the findings to what has been observed with the specific type of insect or animal, then by all means, an internally valid experiment could potentially be designed. The issue, however, arises from the fact that these experiments are barely ever conducted to only study the insect or animal behaviour. Rather, these are used as proxies for human behaviour and therefore, an implicit generalisation of findings to human behaviour is almost invariably given. From that perspective, such experiments score on the extreme low end of the spectrum when the matter of ecological validity is concerned. Therefore, there could be certain level of inferiority attributable to this method when it comes to the issue of external validity. As a result, as long as the questions of intertest can be experimented using human subjects, the choice of animals may not be easy to justify. One can barely think of any topics in modern crowd dynamics research that cannot be experimentally investigated using human subjects. Our research is not concerned with medical interventions or anything of that nature, and for most questions at hand, there should be a way of safely and ethically experimenting our questions with human subjects. Our current experimental capabilities are usually sufficient for almost every relevant topic in crowd research, and the recent records of the variety of experiments appeared in the literature demonstrates that. The significant drop that we have observed in the use of animal and insect methods in crowd research (Haghani, 2020a, b) could also be a testament to the recognition that the notion of experimental validity has implicitly received in recent years from crowd researchers.

## 6. External validity in crowd dynamics experiments
*6.1. Experiment type and its relation to external validity*

The issue of external validity should in general receive more weight when the underlying purpose of our investigation is generalisation and/or prediction as opposed to testing a specific hypothesis or theory. In such cases, it would be essential to minimise artificiality of the design. Consider the experiments exemplified in Figure 5 and contrast them with those shown in Figure 1. None of the three experiments represented in Figure 5 are hypothesis based. In all three, the analyst has certain *assumptions* about what factors that could potentially be important, but beyond that, no hypothesis was made. These exampled experiments are all designed to explore behavioural regularities related to pedestrian exit choice and possibly quantify their effect by fitting a statistical model. In all three cases, the findings of the experiment have been used for *prediction*. In Haghani and Sarvi (2017b) (Figure 5 (1)), experimental observations were used to calibrate an econometric model, and the model was subsequently used as part a broader simulation tool for prediction purposes (Haghani and Sarvi, 2019b). Similarly, Liao et al. (2017) (Figure 5 ( 2)) used the experimental observations to calibrate a regression model and the observations collected from a virtual-reality experiment of Lovreglio et al. (2016) was similarly used for model fitting purposes. These are all models that are developed for prediction purposes (Wagoum et al., 2017), and as such, it is important that the underlying experiment embodies as many factors as possible (as opposed to isolating a single factor) and also not be ecologically too simplistic.

Contrast the experiment of Haghani and Sarvi (2017b) (Figure 5 (1)) to that of Haghani and Sarvi (2019a) (Figure 1 (2)). The former does not make any specific hypothesis about exit choice behaviour and exerts far less experimental control over the exit choice scenarios that participants face. In that case, and compared to experiments in Haghani and Sarvi (2019a), the choice of exit scenarios have been far less "manufactured". As opposed to creating a highly symmetric scenario where the choice has to be made at a certain location where the only influential factor is assumed to be social influence, in the experiments reported in Haghani and Sarvi (2017b), participants could face any choice scenario and make/revise such choices at any point in time. Those scenarios could not be predicted or specified as *a priori*. It was the job of the investigator to explore regularities within that variation of observations through statistical model fitting. While observations of Haghani and Sarvi (2019a) were also analysed through statistical model fitting methods, such model cannot serve as a prediction model, on the exact account of lacking adequate external validity. Such model can only be used for the purpose of hypothesis testing. This was an example experiment where the investigator's aim is to create a setting as naturalistic as possible and in doing so, relinquishes some control over the experiment (in favour of giving more weights to external validity), as opposed to an experiment where the setting is highly manufactured (in order to give more weight to internal validity). Both are legitimate forms of experimental enquiry, assuming that such trade-off is made consciously and in careful consideration of the main purpose of the enquiry.

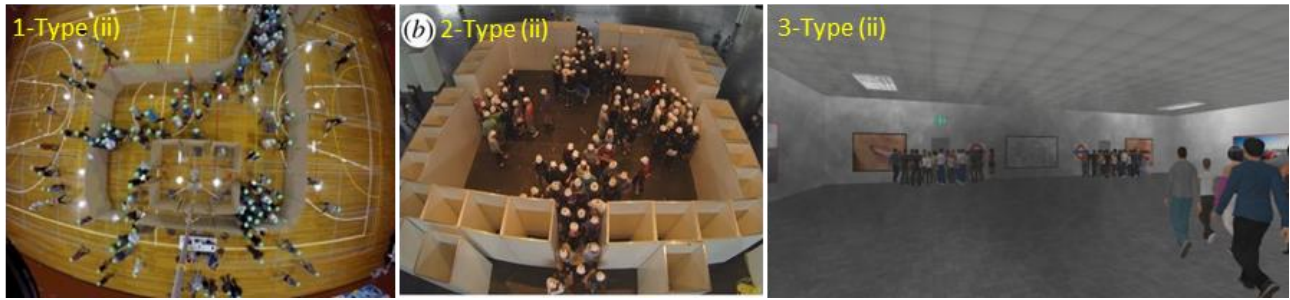

**Figure 5** Examples of non-hypothesis-based crowd experiments where a high degree of ecological realism is essential. (1) Haghani and Sarvi (2017b), (2) Liao et al. (2017), (3) Lovreglio et al. (2016).

*6.2. Dimensions of external validity in crowd dynamics experiments*

In crowd dynamics, *external validity* of an experiment would refer to the extent to which findings of the experiment could be generalisable to real-world crowd and evacuation situations and the extent to which the inferences made are translatable to the actual behaviour of people in real scenarios. Similar to the internal validity, a variety of language and terminologies are commonly used across various fields of science for most of which there is no absolute consensus in definition. For example, the notion of external validity could also be considered inextricably linked with *criterion validity* (or criterion-related validity) concerning whether the results of a method correspond with an external criterion (Telser and Zweifel, 2007). This, in crowd dynamics, could for example mean whether an evacuation model that we have developed or calibrated using experimental data could predict behaviour of people observed in CCTV footage of a real emergency (i.e., an external criterion). This is sometimes also referred to as *predictive validity*[2] (Kealy et al., 1990).

Similar to internal validity, a number of different dimensions can be recognised in relation to external validity of crowd experiments. If the findings of a crowd experiment are found reproducible through other experiments, then experimental findings possess *convergent validity*. This is separate to criterion validity, in that, the benchmark of comparisons in this case need not be an external (i.e., non-experimental) criterion, rather, it could just be another experiment. For example, in experiments of Haghani and Sarvi (2017b), the investigators had a specific physical setup for their exit choice (Figure 6 (2)). This physical setup was inspired by the setup presented in a previous survey (Figure 6 (1)). Both experiments had essentially produced similar models when compared based on the statistical model fitted based on their respective observations. In a new set of experiments reported in Haghani et al. (2020) a different physical setup was tested and similar modelling findings were again observable (Figure 6 (3)). This is an example of a case of testing convergent validity in crowd experiments, as a dimension of external validity (Haghani and Sarvi, 2019b).

If the experimental setup (particularly, the physical setup) and the treatments bear enough resemblance to real-world crowd and evacuation situations, then the crowd experiment scores higher on *ecological validity*. Achieving this criterion requires that simplifications be kept to a minimum during the design process and that the experiment be designed as realistic as possible. The notion of ecological validity is clearly relative. No experiment can have perfect ecological validity and a single experiment in isolation cannot be readily evaluated on this dimension. It is through comparisons with similar experiments on similar topics that one can claim an experiment to be more/less ecologically valid.

If findings of a crowd experiment are found reproducible by other populations (e.g., across cultures, age groups etc) then the findings of the experiment demonstrate *population validity* (see Lin et al. (2020) as an example). Another dimension is *temporal validity* which concerns whether findings of crowd experiment observed at one particular point in time are going to remain valid over time. These dimensions (criterion, convergent, ecological, population, and temporal validity) can all be considered manifestations by which external validity of an experiment can be assessed and tested. This often needs to be done through accumulation of evidence and subsequent testing and cannot necessarily be accommodated all within a single experimental study.

---

[2] Often, the difference between these three concepts is so subtle that any effort of further differentiating them may add confusion and no value. Therefore, we may as well consider these terms interchangeably usable in crowd research.

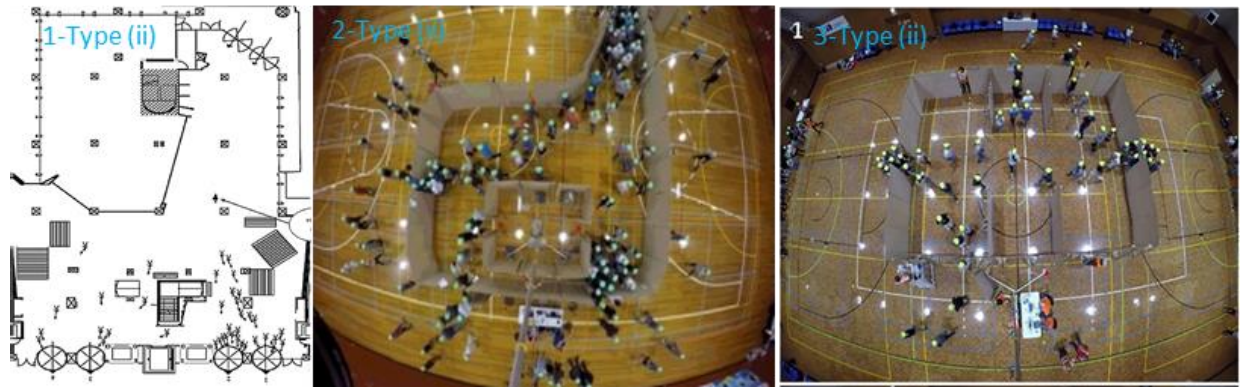

**Figure 6** Examples of a series exploratory Type (ii) experiments where the dimension of convergent validity was tested. (1) & (2) Haghani and Sarvi (2017b), (3) Haghani et al. (2020).

*6.3. Threats to external validity of crowd dynamics experiments*

Similar to internal validity, a range of threats are identifiable that can potentially affect external validity of crowd experiments. For example, crowd experiments using samples that are not representative of the general population are at the threat of *sampling bias* to their external validity. This perhaps pertains to the majority of crowd experiments in which participants are university students. The extent to which the issue of sampling bias affects external validity of these experiments, however, needs to be assessed on a case-by-case basis and in consideration of the topic at hand. There may be cases where sampling bias is not necessarily a significant threat to the external validity of crowd experiments.

Crowd experiments conducted in highly simplistic settings face the threat of *setting artificiality and situational effects* to their external validity. This could be any of the hypothesis-based experiments that were exemplified in Figure 1. When the main aim of the experiment is testing a hypothesis or theory, the experimenter may choose to accept artificiality in order to gain higher levels of internal validity. Also, the fact that we test our questions in one single layout and wish to generalise across all possible architectural layouts is another dimension of this threat to external validity, i.e., artificiality. Therefore, to the extent possible, the investigator may choose to incorporate a variety of settings in the experiment in order to test situational effects. Considering that time and resources are always limited in conducting crowd experiments, this may not necessarily be possible. For example, after conducting crowd exit choice experiments reported in Haghani and Sarvi (2017b), it took the investigators another four years to be able to test the said exit choice experiments in a different physical layout and with a different population (see Haghani et al. (2020) and Figure 6).

Let us assume that we have designed and scheduled a crowd evacuation experiment. A few days before the experiment, a major crowd crush incident happens in our city exposing most of our participants to its media coverage. This may affect the behaviour of those participants during the experiment, making our sample a very specific population at that time, thereby posing the threat of *history* to external validity.

Consider an evacuation experiment that aims to mimic an acute sense of emergency in a crowd, such as a terror attack. Participants who may engage in pushing other people out of their way in fear of their life when facing such threatening scenario in real life, may refrain from showing such socially unacceptable behaviour in our experiment. This tendency of participants to modify their behaviour because they are being watched by the researchers constitutes the threat of *Hawthorn effect* to external validity. Another example could be an experiment when participants realised (as a result of the investigators accidentally sharing the information with them during the brief or through participants inferring from other cues) that we are interested in their decision changing behaviour during evacuations (Haghani et al., 2020). This may prompt them to change their decision more often than they naturally may do in order to please the investigator and give them the data that they need. This is also sometimes referred to as *experimenter/investigator bias* when participants modify their behaviour in order to give us the desired outcome, by trying to infer the underlying purpose of the experiment. This could compromise external validity of the observations to some degrees and constitutes a good reason to mask the underlying questions of the experiment to the extent possible in order to minimise this effect.

Consider a crowd experiment that entails a social, moral or altruistic component. Examples could be seen in the work of [Bode et al. (2015b)](#) or [Moussaïd and Trauernicht (2016)](#)). If participants get to learn or infer the underlying purpose of the experiment, or if the investigator inadvertently communicates that through subtle cues, then that may prompt participants to try to demonstrate behaviour that makes them look socially more desirable (here, more altruistic behaviour for example). This is often labelled as *social desirability* or *warm glow* effect ([Hartmann et al., 2017](#); [Lopez-Becerra and Alcon, 2021](#)) and again can be mitigated by ensuring that participants do not get to learn the underlying question of the study. This sometimes ma require us to resort to the deception technique, for example, when briefing the participants. But ethical considerations will also require us to give them a debrief after the experiment is completed and tell them about any possible deception.

Similar to the effect mentioned above, consider a between-group design where we are interested in determining effectiveness of a certain behavioural modification on evacuation performance. Two groups are performing the tasks in parallel[3]; one is the control group, and one is the group that were instruct/train to adopt the said behavioural modification. We take every possible measure to ensure that the two groups do not communicate or interact with each other (during breaks etc). Still, members of the control group realise that they are the control group, and the other group is given certain instructions. This might prompt them to modify their behaviour in some ways to make their evacuation behaviour more efficient (even if they have no idea what the intervention for the treatment group is). This phenomenon is referred to as *John Henry effect* which is another threat to external validity[4].

## 7. Discussions and concluding remarks

The aim of this work was to bring the issue of experimental validity closer to the forefront of discussions in crowd dynamics, an issue that is going to become increasingly relevant as more crowd experiments are conducted and reported. These discussions aimed to provide an initial benchmark that investigators can use to ensure they address relevant validity-related dimensions in their design and reporting of their experiments and to also provide a formulated basis on which validity of such experiments can be assessed. The research environment where we conduct crowd experiments is typically a very controlled and artificial one, unlike the everyday world to which we wish to generalise our findings. While it is important for crowd experiments to be able to control for all possible extraneous variables to be sure of the casual relationships that are discovered, this often comes at the cost of external validity to a world in which variables of interest almost invariably occur in the presence of other extraneous variables. Questions such as when it is acceptable or unjustifiable to introduce an artificial design and what are the criteria for making such determination were the main motivations of this work.

---

[3] It may be worth mentioning that some of these threats to validity are specific to between-subject experiments only (while some others are also specific to within-subject designs). Virtual-reality experiments could potentially be designed using either approach. But when it comes to field-type experiments, the majority of experiments have, thus far, been between-subject designs, where we expose the same group of people to multiple treatments and several repetitions. This could be generally an issue of resources. In field-type experiments, we rely on concurrent presence of a large group of participants in our laboratory location, as without their concurrent present there will be no crowd to conduct the experiment with. This may be one of the unique characteristics of crowd experiments. Due to limitations in our resources and the number of participants that we can recruit, we often cannot split them into more than one group. Therefore, in the majority of field-type experiments, the design has been within subject.

[4] While we have in this work separated a list of threats to internal and external validity of crowd experiments, it should be noted that the belonging of some of these items to internal versus external validity is not very clear cut and is fuzzy and perhaps arguable. This arises from the fact that the distinction between internal and external validity itself is not perfect. They are interrelated, in that, without internal validity (i.e., where there is no genuine causal effect), there would be nothing to generalise (hence, the discussion around external validity will be meaningless). Therefore, for the purpose of crowd experiment design, the distinction between whether these threats exactly affect internal validity or external validity may not even be necessary. Threats to internal validity are factors that ultimately jeopardise external validity of the experiment too. It is essential that during the design and execution phase of any crowd experiment, any potential threat to validity (regardless of internal or external) be identified and efforts be made to mitigate them.

Another factor that motivated the current work was the uniqueness of experiments reported in crowd dynamics as well as the fact that experimental work in this domain is relatively young. Sometimes, we conduct experiments that have clear treatment conditions or interventions (Bode and Codling, 2019; Bode and Codling, 2013; Garcimartín et al., 2016; Zuriguel et al., 2020) resembling experiments in clinical medicine. Sometimes we perform experiments that do not entail any particular intervention and are simply exploratory resembling some field-type experiments in behavioural economics (Levitt and List, 2009; Reiley and List, 2007; Viceisza, 2016). This is just a reflection of the multidisciplinary nature of this field (Haghani, 2021). In clinical medicine for examples, trials and experiments are far more established than crowd dynamics. They often have similar structure to one another, and as a result, clear protocols have been developed to assess their validity (Slack and Draugalis Jr, 2001). The uniqueness of most crowd experiments and the fact that different studies adopt different experimental methodologies have, thus far. made it impossible for the crowd community to develop such universal protocols. While commonalities and categories are identifiable within the experimental literature of crowd dynamics, the diversity of topics and methodologies have made the protocols of good practice in experiment design difficult to establish.

The cornerstone of the discussions presented in this work was that a major factor associated with the assessment of validity in crowd experiments is determining the category with which our experiment identifies. This categorisation should be made on the basis of the main purpose of the experiment (and the experimental method). It is important, for the sake of validity assessment (for both investigators and reviewers), that the study identifies itself clearly as one of the main categories. While experiments of crowd dynamics could be used for dual or multiple purposes, there is always one main purpose the drives the investigation and that factor should be made clear from the conception phase through to the reporting and publication stage. From this perspective, it is crucial that we determine as *a priori* whether our experiment is designed to test[5] a specific hypothesis or theory[6], whether it is a largely exploratory study designed to establish some empirical behaviour regularities through statistical model fitting or model calibration, whether we are looking at the effect of a certain behavioural or architectural intervention on crowd performances, or whether we are looking at the partial calibration of a specific subset of parameters within a pre-existing model. Each of these categories, as we demonstrated through existing practice in the literature may require their own validity considerations. Some may justify high levels of simplicity and compromise of ecological (and thereby, external) validity and some cannot be justified unless designed realistically enough. The discussions presented in the previous section have been summarised in an abstract format in Figure 7.

---

[5] A behavioural theory is always provisional, in the sense that it is only a hypothesis. We can never prove it no matter how many times the result of experiments agrees with some theory, and we can never be sure that the next time the results will not contradict the theory. On the other hand, we can disprove a theory by finding even a single observation that disagrees with the theoretical prediction. Each time new experiments have observed to agree with the predictions, the theory survives, and our confidence in it is increased. But if an observation is found to disagree, we must abandon or modify the theory. For more discussions on this I refer to Hawking, S., 2006. The theory of everything. Jaico Publishing House.).

[6] A theory is just a model of behaviour or a restricted part of it and a set of rules that connects quantities in the model to observations that we make. A good theory must satisfy two requirements: it must accurately describe a large class of observations, and it must be able to make predictions about the results of future observations. We do not have a general crowd theory, maybe an array of small theories each of which capable of explaining parts of crowd-related phenomena. We also do not have many original theories of crowd behaviour or crowd movement. Most of our theories are derived theories or borrowed or modified theories. We break our problems down to a number of partial theories, as a universal theory is hard to achieve. Each of these partial theories describes and predicts a certain limited class of observations, neglecting the effects of other variables. But if everything depends on everything else, it would be very hard to be able to describe any part of behaviour in isolation. Nevertheless, it is certainly the way that we have made progress in this field in the past.

One clear implication of the earlier discussion is that they warn us against conducting experiments for the sake of conducting experiments without having clear questions and investigation aims in mind. It is always possible to recruit a group of participants and create a crowd in a laboratory setting and have them perform certain tasks (of walking, evacuation etc). No matter what the task is, if we dig deep enough, there are always some regularities and patterns that can be found through these experiments. But this is where the issue of experimental validity confronts us. Do these behavioural regularities observed in our laboratory setting mean anything for the real-world? To be more precise, if, for example, we are not testing a theory/hypothesis or are not trying to calibrate/measure a very specific parameter/quantity, then conducting an experiment in an outlandish highly simplified or highly specific architectural setting would not add much value. Yes, we may be able to draw some inferences, but then, we face the question from those who assess our work: what does this means for real-world crowded situations? A question that we should be able to defend. Experiments always entail a certain level of compromise of realism, an attribute that cannot be completely avoided but can be mitigated. If an investigation can be carried out using field data in real environments, then it is not justified to take such investigation to an artificial laboratory setting. Practice of this sort could potentially diminish the value of experimental work in crowd dynamics.

On the other hand, having a clear and robust conversation within the crowd dynamics community about when investigations should be conducted using field data ([Corbetta et al., 2018](); [Pouw et al., 2020]()) and when experimentation (and the compromise of realism) is justified (i.e., whether the gain of internal validity justifies the sacrifice of external validity) would help us better defend our engagement in experimental work. When faced with the question that why we use artificial experiments to develop/calibrate an exit choice model of evacuation when there is CCTV footage of real evacuations available, we can produce an argument grounded in the notion of validity. When calibrating a model an important element is adequate variability within the set of observations. Another important factor is the accuracy of measurements. Neither of these attributes are necessarily offered by a few seconds of CCTV footage of an earthquake evacuation, for example. It might be impossible, for certain questions in crowd dynamics, to establish a cause-and-effect relationship without exerting experimental control. That is why we often choose to take these enquiries to the laboratory and create an artificial evacuation scenario and test them experimentally (knowing that it may have significant contextual differences with a real-world life-threatening situation). This enables us to create the required quantity of observations and the degree of variability (within those observations) that we desire and to measure our observations accurately. This is comparable to what a road safety researcher might do to, for example, understand the impact of recreational drug use on drivers' performance. Once we achieve a model, however, we are aware that the performance of such model can be assessed using an external criterion (here, observations obtained from the CCTV footage) in order to assess the degree of validity. Here, we labelled such practice a test of criterion validity as part of the broader concept of external validity. Questions may be similarly asked regarding why we often perform our experiments in a small temporary and artificial psychical setup instead of a real building. The answer would be similar to the previous question. If we seek model development/calibration from our investigation, then making alternations to the physical environment as well as accurate measurements are essential, and these may not necessarily be feasible to achieve in an evacuation drill experiment in a real-size building.

Another dimension that was brought up as part of the external validity discussions was the component that we referred to convergent validity, concerning whether an experimental finding is observable through other experiments. As mentioned earlier, it takes only one (internally valid) experiment to debunk a theory but if we seek to ascertain the validity of a theory or finding, then that requires accumulation of empirical evidence. Replications of experiments are often acceptable practice in fields such as medical sciences or psychology. The popularity of meta-analytical studies in those fields of sciences are testaments to the fact that similar/comparable experiments have been reported on certain questions. Take the topic of the effect of acute alcohol consumption on driver performance as an example. Dozens of independent papers have reported on experimentation of this topic throughout the years (experiments that may have slight differences but essentially look at the same underlying question) and this has been reported and summarised in relevant meta-analytical studies (Irwin et al., 2017). To this date, there has not been a single study in crowd dynamics that can be labelled as a classic meta-analysis (the only exception is the very recent work of Hu and Bode (2021) as a meta-analysis of the effect of social groups on evacuation time, published as this manuscript was being drafted). This is certainly not for the lack of studying same topics over and over. Take the simple topic of faster-is-slower as an example that has been investigated several times in previous experimental work. This lack of meta-analytical findings could, however, be due to the fact that independent experiments report their findings in ways that are not comparable. This could be partly because previous studies might have used vastly different and incomparable methods (e.g., experiments with ants, mice and humans). Another potential reason is that, in order for new studies to be publishable, they need to present elements of novelty, and for such considerations, investigators may choose to carry out their experiment slightly different to the previous ones or analyse the results in different ways, rendering the comparisons across studies rather impossible. This issue has so far hindered the ability of crowd researchers to draw unified conclusions even on their most fundamental questions or to test convergent validity of most of the experiments. A solution that comes to mind is that instead of pure replication of previous experimental findings, we may resort to *constructive replications* (i.e., deliberately avoiding duplication of procedures by testing a preposition independent of the method that is applied for testing) (Johnson and Bouchard, 2005). For example, in an experiment reported in Haghani et al. (2020), as part of the fining, the investigators observed that presenting participants with monetary incentives had a significant effect on evacuation efficiency and reduced evacuation time. In a later study, Xue et al. (2021) adopted this method and made a similar observation while adding the component of "limited visibility" to their design. While this provided evidence of convergent validity for the said finding, it also avoided the issue of complete duplication by providing an additional element in the experiment design, thereby, maintaining the novelty.

Another matter that should be highlighted is that when conducting external (and more specifically criterion validity) testing, comparisons need to be made with a more valid construct than the experimental setting itself (e.g., validating an empirical model against a real CCTV footage). Often it is observed, in crowd dynamics studies, that numerical simulation findings are used as the benchmark of validation of empirical observations. For instance, one might come across statements claiming that "our simulation model produced patterns that confirm the herd behaviour assumption". Numerical simulation models are the artefact of the behavioural assumptions that we feed them (directly or indirectly through our formulations). As a result, a purely theoretical numerical model cannot be used as the benchmark of validity testing for experimental findings.

While our previous discussions were exclusively confined to the issue of validity in crowd experiment design, similar questions can also be raised in relation to the validity of statistical inference in crowd experiments. This issue is outside the scope of this work and readers are referred to the work of Bode and Ronchi (2019) for aspects related to that matter. Matters of validity in statistical modelling are those that essentially pertain to post-experiment activities and the analysis, i.e., given an already executed experiment and a dataset, how we can ensure that our inferences possess the highest possible degree of validity. Discussions of the current study mainly concerned the issue of validity as a logical issue relating to the design and execution of the experiment (consistent with Campbell and Stanley (2015)) rather than a statistical issue.

Despite clear limitations in validity and generalisability, crowd experiments are going to continue and perhaps will remain our best (and sometimes the only) tool for many of investigations in this area. It is through the complementary role of experimental and field studies that empirical knowledge in crowd dynamics can accumulate. Neither of these approaches supersedes the other when it comes to the issue of validity. While field observations may be more realistic and give the impression of higher external validity, investigations conducted in experimental settings may have stronger elements of internal validity. If we have a specific hypothesis or theory in mind, it is very rare that any field data (collected in a completely inobtrusive way and free of any experimental intervention) gives us enough information to test that hypothesis[7]. This trade-off should always be considered in empirical investigations in crowd dynamics[8]. As mentioned earlier, without internal validity, any discussion on external validity of a set of observations will essentially be pointless. Finally, our discussions suggest that any of the three established experimental methods in crowd dynamics could potentially produce valid results. And similar to the relation between field and experimental data, there is also no superiority between virtual-reality, field-type and drill experiments when it comes to the issue of validity. When designed properly, each methodology could potentially produce valid results. Although, without suggesting this as a universal rule, in majority of topics, virtual-reality and field-type crowd experiments may present stronger elements of controllability, while drill experiments may present stronger elements of realism and hence ecological (thereby, external) validity. Between the virtual-reality and field-type experiments, often the former could offer more control to the investigator over the cause-and-effect relations. However, this is not the only consideration to make when deciding the experimental method. The topic at hand often itself dictates the most suitable method. For instance, if the investigation is all about the physical aspects of pedestrian movements, then a virtual-reality setting may not be most suitable choice. But when it comes to matters of pedestrian decision-making, for example, then there would be a closer trade-off when choosing which method to employ.

I believe that this work could be followed up by developing specific protocols of good practice for the design and execution of crowd experiments in relation to each of the three main experimental methods in this field. This could be along the lines of discussions provided by Gwynne et al. (2019) and Gwynne et al. (2020) in relation to evacuation drills. Such protocols of good practice are currently missing for virtual-reality and field-type crowd experiments. Also, reiterating a point mentioned earlier, validity of experimental findings pertains both phases of design/execution and statistical analysis. From that perspective, we should not lose sight of the fact that drawing valid inferences necessitates that good practice in design/execution is followed by good practice in statistical modelling (Bode and Ronchi, 2019).

---

[7] Field data in crowd research can however be good for forming new hypotheses and theories so we can test them later in experimental situations. Field research in crowd dynamics can essentially feed into the experimental research and vice versa.

[8] A potential compromise between desired features of experimental and field research in crowd dynamics could be *quasi experiments*, where we can exert certain degrees of intervention in the real world without recruiting participants and without people knowing that they are being observed by experimenters. This potential method did not warrant much coverage in our discussions mainly because there is not much precedent of this approach in crowd research yet. An example could be conducting a nudging experiment (through flashing lights, for example) in a pedestrian crossing location or train station and observing the change of behaviour. Clearly, applicability of such approach would be limited to very specific types of enquiries in crowd research.

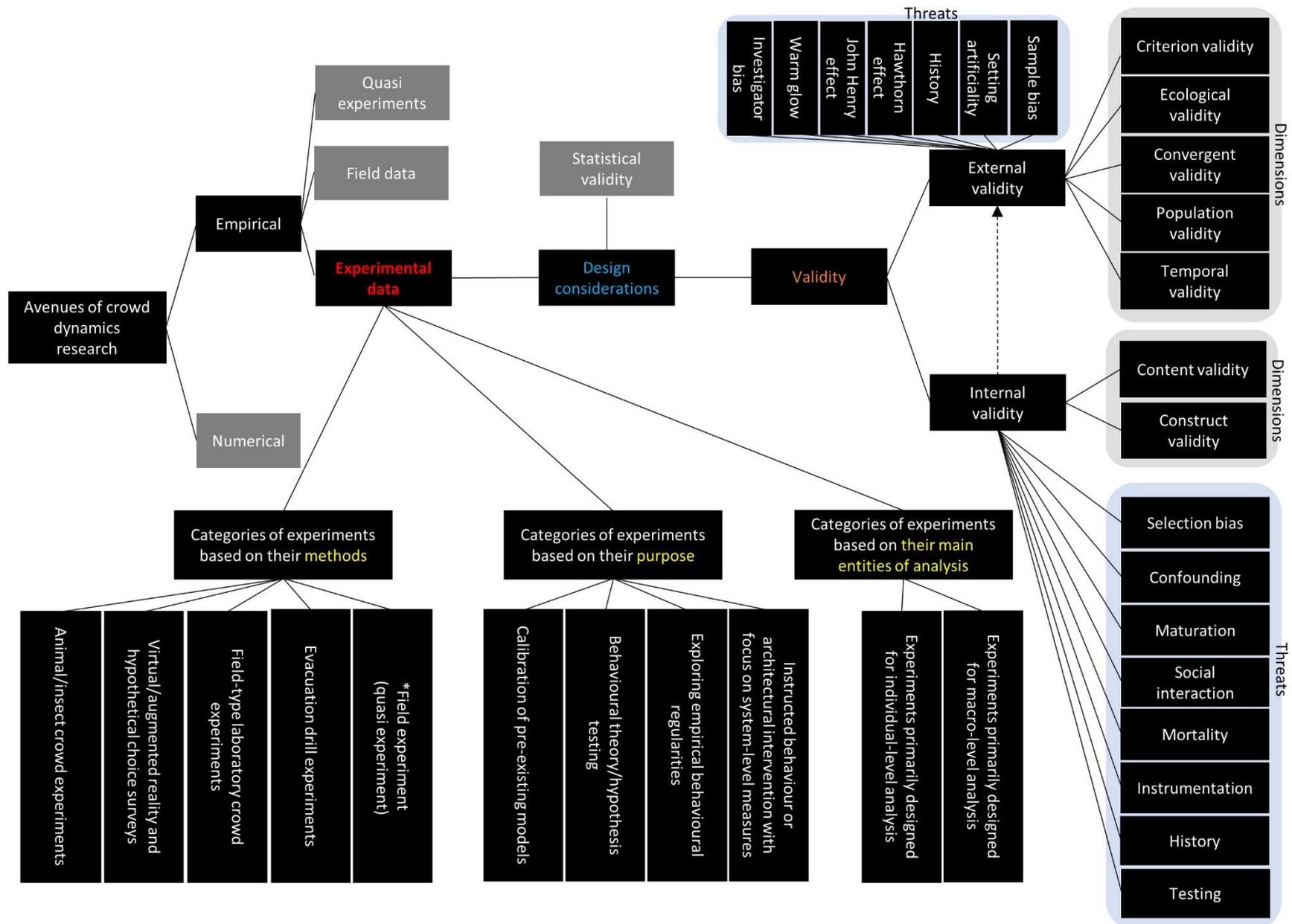

**Figure 7** Dimensions of validity and threats to validity in experimental crowd dynamics.

**Acknowledmengts**

This research was funded by Australian Research Council grant DE210100440.
**References**

Bishop, R.C., Boyle, K.J., 2019. Reliability and validity in nonmarket valuation. Environmental and Resource Economics 72, 559-582.

Bode, N.W., Codling, E.A., 2019. Exploring determinants of pre-movement delays in a virtual crowd evacuation experiment. Fire technology 55, 595-615.

Bode, N.W., Holl, S., Mehner, W., Seyfried, A., 2015a. Disentangling the impact of social groups on response times and movement dynamics in evacuations. PloS one 10, e0121227.

Bode, N.W., Miller, J., O'Gorman, R., Codling, E.A., 2015b. Increased costs reduce reciprocal helping behaviour of humans in a virtual evacuation experiment. Scientific reports 5, 1-13.

Bode, N.W., Ronchi, E., 2019. Statistical model fitting and model selection in pedestrian dynamics research. Collective Dynamics 4, 1-32.

Bode, N.W.F., Codling, E.A., 2013. Human exit route choice in virtual crowd evacuations. Animal Behaviour 86, 347-358.

Burstedde, C., Klauck, K., Schadschneider, A., Zittartz, J., 2001. Simulation of pedestrian dynamics using a two-dimensional cellular automaton. Physica A: Statistical Mechanics and its Applications 295, 507-525.

Campbell, D.T., 1986. Relabeling internal and external validity for applied social scientists. New Directions for Program Evaluation 1986, 67-77.

Campbell, D.T., Stanley, J.C., 2015. Experimental and quasi-experimental designs for research. Ravenio Books.

Cook, T.D., Campbell, D.T., 1979. The design and conduct of true experiments and quasi-experiments in field settings, Reproduced in part in Research in Organizations: Issues and Controversies. Goodyear Publishing Company.

Corbetta, A., Meeusen, J.A., Lee, C.-m., Benzi, R., Toschi, F., 2018. Physics-based modeling and data representation of pairwise interactions among pedestrians. Physical review E 98, 062310.

Dias, C., Lovreglio, R., 2018. Calibrating cellular automaton models for pedestrians walking through corners. Physics Letters A 382, 1255-1261.

Drury, J., Brown, R., González, R., Miranda, D., 2016. Emergent social identity and observing social support predict social support provided by survivors in a disaster: Solidarity in the 2010 Chile earthquake. European Journal of Social Psychology 46, 209-223.

Drury, J., Cocking, C., Reicher, S., 2009a. Everyone for themselves? A comparative study of crowd solidarity among emergency survivors. British Journal of Social Psychology 48, 487-506.

Drury, J., Cocking, C., Reicher, S., 2009b. The nature of collective resilience: Survivor reactions to the 2005 London bombings. International Journal of Mass Emergencies and Disasters 27, 66-95.

Garcimartín, Á., Parisi, D.R., Pastor, J.M., Martín-Gómez, C., Zuriguel, I., 2016. Flow of pedestrians through narrow doors with different competitiveness. Journal of Statistical Mechanics: Theory and Experiment 2016, 043402.

Guo, N., Ding, Z.-J., Zhu, K.-J., Ding, J.-X., 2020. Characteristics of pedestrian flow based on an improved least-effort model considering body rotation. Journal of Statistical Mechanics: Theory and Experiment 2020, 073401.

Gwynne, S., Amos, M., Kinateder, M., Bénichou, N., Boyce, K., Natalie van der Wal, C., Ronchi, E., 2020. The future of evacuation drills: Assessing and enhancing evacuee performance. Safety Science 129, 104767.

Gwynne, S.M.V., Kuligowski, E.D., Boyce, K.E., Nilsson, D., Robbins, A.P., Lovreglio, R., Thomas, J.R., Roy-Poirier, A., 2019. Enhancing egress drills: Preparation and assessment of evacuee performance. Fire and Materials 43, 613-631.

Haghani, M., 2020a. Empirical methods in pedestrian, crowd and evacuation dynamics: Part I. Experimental methods and emerging topics. Safety Science 129, 104743.

Haghani, M., 2020b. Empirical methods in pedestrian, crowd and evacuation dynamics: Part II. Field methods and controversial topics. Safety Science 129, 104760.

Haghani, M., 2020c. Optimising crowd evacuations: Mathematical, architectural and behavioural approaches. Safety Science 128, 104745.

Haghani, M., 2021. The knowledge domain of crowd dynamics: Anatomy of the field, pioneering studies, temporal trends, influential entities and outside-domain impact. Physica A: Statistical Mechanics and its Applications 580, 126145.

Haghani, M., Bliemer, M.C.J., Rose, J.M., Oppewal, H., Lancsar, E., 2021. Hypothetical bias in stated choice experiments: Part II. Conceptualisation of external validity, sources and explanations of bias and effectiveness of mitigation methods. Journal of Choice Modelling 41, 100322.

Haghani, M., Sarvi, M., 2017a. Following the crowd or avoiding it? Empirical investigation of imitative behaviour in emergency escape of human crowds. Animal Behaviour 124, 47-56.
23